\documentclass[english,aps,prl,superscriptaddress,twocolumn,showpacs,footinbib]{revtex4-1}
\usepackage{graphicx}  % needed for figures
\usepackage{dcolumn}   % needed for some tables
\usepackage{bm}        % for math
\usepackage{amssymb}   % for math
\usepackage{siunitx}   % for math
\usepackage{babel}
\usepackage{verbatim}
\usepackage{amsmath}
\usepackage{setspace}
 \usepackage{epstopdf}
\graphicspath{{figures/}}
\usepackage{siunitx}
\usepackage{float}
\usepackage{xcolor}
\usepackage{soul}
\begin{document}

 %The following information is for internal review, please remove them for submission
\widetext

\title{Quantitative vectorial magnetic imaging of multi domain rock forming minerals using nitrogen-vacancy centers in diamond}

\author{E. Farchi}
\affiliation{Dept. of Applied Physics, Rachel and Selim School of Engineering, Hebrew University, Jerusalem 9190401, Israel}

\author{Y. Ebert}
\affiliation{The Institute of Earth Sciences, The Hebrew University of Jerusalem, Jerusalem 91904, Israel}

\author{D. Farfurnik}
\affiliation{Racah Institute of Physics, The Center for Nanoscience and Nanotechnology, The Hebrew University of Jerusalem, Jerusalem 9190401, Israel}

\author{G. Haim}
\affiliation{Dept. of Applied Physics, Rachel and Selim School of Engineering, Hebrew University, Jerusalem 9190401, Israel}

\author{R. Shaar}
\affiliation{The Institute of Earth Sciences, The Hebrew University of Jerusalem, Jerusalem 91904, Israel}

\author{N. Bar-Gill}
\affiliation{Racah Institute of Physics, The Center for Nanoscience and Nanotechnology, The Hebrew University of Jerusalem, Jerusalem 9190401, Israel}
\affiliation{Dept. of Applied Physics, Rachel and Selim School of Engineering, Hebrew University, Jerusalem 9190401, Israel}

\date{\today}

\begin{abstract}
Magnetization in rock samples is crucial for paleomagnetometry research, as it harbors valuable geological information on long term processes, such as tectonic movements and the formation of oceans and continents. Nevertheless, current techniques are limited in their ability to measure high spatial resolution and high-sensitivity quantitative vectorial magnetic signatures from individual minerals and micrometer scale samples. As a result, our understanding of bulk rock magnetization is limited, specifically for the case of multi-domain minerals. In this work we use a newly developed nitrogen-vacancy magnetic microscope, capable of quantitative vectorial magnetic imaging with optical resolution. We demonstrate direct imaging of the vectorial magnetic field of a single, multi-domain dendritic magnetite, as well as the measurement and calculation of the weak magnetic moments of an individual grain on the micron scale. These results pave the way for future applications in paleomagnetometry, and for the fundamental understanding of magnetization in multi-domain samples.
\end{abstract}

%\pacs{76.30.Mi}
\maketitle

%%%%%%%%%%%%%%%%%%%%%%%%%%%%%%%%%%%%%%%%%%%%%%%%%%%%%%
%% INTRODUCTION
%%%%%%%%%%%%%%%%%%%%%%%%%%%%%%%%%%%%%%%%%%%%%%%%%%%%%%

\paragraph{}% Paleomagnetism introduction
When igneous rocks in Earth's crust are formed by cooling of hot magma they acquire thermoremanent magnetization (TRM) parallel and proportional to the ambient field at the time of cooling. TRM can remain stable for millions of years preserving valuable geological magnetic information on complex long term processes, such as plate tectonic movements and formations of oceans and continents \cite{Tauxe2016}. Despite its importance, little is known on the details of how thermoremanent magnetic information is acquired and retained in rocks \cite{dunlop1997}. Paleomagnetism, the science of studying natural magnetic information, heavily relies on N\'{e}el theory \cite{Neel1955} of single-domain (SD) minerals. Yet, the dimensions of rock-forming minerals typically exceed the sub-micrometric threshold size for SD. As such, TRM is mostly held by multi-domain (MD) particles rather than by SD. In the absence of a general analytical formulation for TRM in MD, there is a growing need for direct observations of natural MD magnetization. Particularly, vectorial imaging of magnetic fields generated by MD particles with sub-micron spatial resolution and micrometer scale sample-detector distance, along with high sensitivity ($\mu T/\sqrt{Hz}$), is critical for understanding the mechanism controlling the geometry and size of MD arrangements, the total moment exerted by individual natural crystals, and the stability of MD magnetic information over geological times. 

\paragraph{} % Other measurement methods 
There are a number of magnetic imaging techniques that have been used in paleomagnetic research in a complementary fashion: Kerr effect \cite{Hoffman1987} , Magnetic Force Microscopy (MFM) \cite{Pokhil1996}, electron holography \cite{Harrison2002}, and SQUID microscopy \cite{Weiss2007}. Recently, a newly developed method based on nitrogen-vacancy (NV) magnetic microscopy \cite{Simpson20016,Pelliccione2014,Tetienne2014}, has been demonstrated by \cite{Fu2014,Fu2017} in the context of paleomagnetometry. NV magnetic microscopy enables direct measurements of the three components of the magnetic field vector at a constant height above the sample in a room temperature environment. It features sub-micrometer spatial resolution of quantitative data with relatively high sensitivity and simple operation.

\paragraph{} % NV in general introduction 

%The NV center is composed of a substitutional nitrogen atom (N) and a vacancy (V) on adjacent lattice sites in the diamond crystal.

\begin{center} 
\begin{figure}[tbh]	
	\includegraphics[width=1\columnwidth]{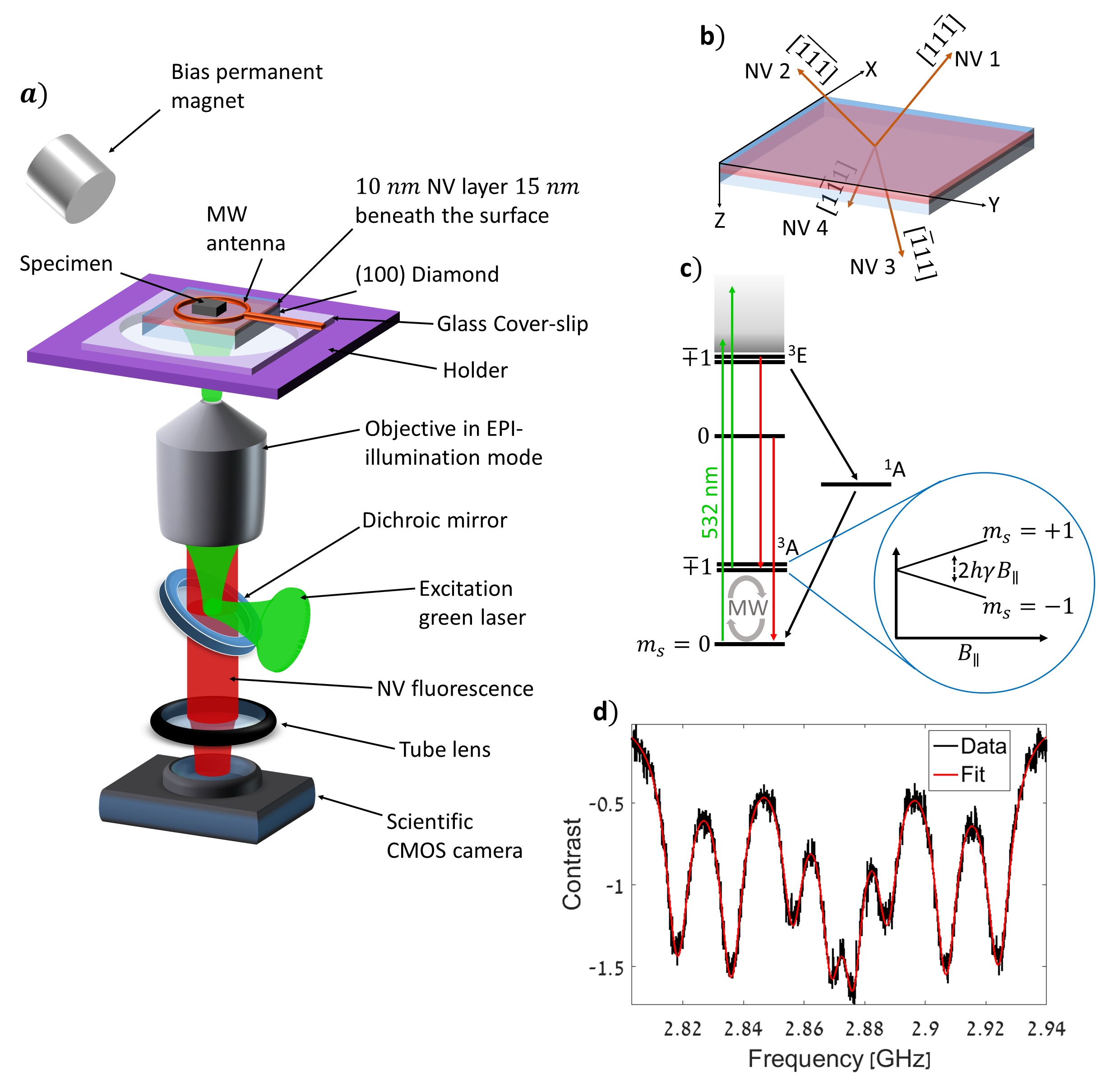} %fig3a.eps
	\caption{Schematic of the experiment, the NV structure and measurement process. (a) Custom-built Wide-field magnetic imaging microscope. The rock sample is placed on the surface of a diamond chip, which is implanted with a high density thin layer of nitrogen-vacancy (NV) centers near the surface. The diamond is attached to the cover-slip using immersion oil which is glued to the holder. Optical pumping green laser is incident through the bottom-polished side of the diamond surface using the objective in EPI mode. Coherent MW-field manipulation, which is created by an MW antenna, is located near the diamond surface containing the NV layer. The NV fluorescence passes through the diamond, the cover-slip and the dichroic mirror, and is then imaged onto the sCMOS camera using the objective and a tube lens. (b) Schematic of sample orientations showing the $ (100) $ diamond cheep (blue) containing the NV layer (red) and the four NV orientations (orange arrows). Crystallographic directions of the diamond are depicted as well as the Cartesian coordinate system. (c), Energy-level diagram of the NV center. An external static magnetic field along the NV symmetry axis Zeeman-shifts the energy of the $m_s=\pm 1$ levels linearly (magnified on the right). (d), Full optically detected magnetic resonance (ODMR) spectra for one pixel. The characteristic dips correspond to the resonance transitions between the energy levels of the NV center's spin. Multiple resonances are observed, corresponding to the different NV orientations in the diamond crystal.}
    \label{fig: systemSchematic}
\end{figure}
\end{center}

The nitrogen-vacancy (NV) center is a point defect in the diamond's lattice with C3V symmetry. It consists of a substitutional nitrogen atom (N) adjacent to a carbon vacancy (V) that is oriented along one of the four crystalline directions ($ [11\overline{1}], [\overline{1} \overline{1} \overline{1}], [\overline{1} 1 1] $ and $ [1 \overline{1}1] $) (Fig. \ref{fig: systemSchematic} (b)). The electronic structure of the negatively charged NV center has a spin-triplet ground state, where the $m_s=\pm 1$ sublevels experience a zero-field splitting ($\sim 2.87$ \si{GHz}) from the $m_s = 0$ sublevel due to spin-spin interactions. When an external static magnetic field is applied along the NV symmetry axis, the degeneracy of the $m_s=\pm1$ energy levels is lifted linearly via Zeeman splitting (Fig. \ref{fig: systemSchematic} (c)) \cite{NVinDiamond}.
% Application of an external static magnetic field along the NV symmetry axis Zeeman shifts the energy of the $m_s=\pm 1$ levels linearly.
The NV spin state can be initialized in the $m_s= 0$ state with off-resonant laser excitation, coherently manipulated with resonant microwave (MW) pulses, and read out optically via spin-state-dependent fluorescence intensity of the phonon sideband \cite{Taylor2008}.

Here, we explore the potential of NV magnetic microscopy in observing magnetic domains and measuring the effective dipole moment of large multi-domain arrays. We present two case studies, depicting direct quantitative, vectorial magnetic imaging of a single dendrite in the multi-domain regime, and of a pair of magnetite dendrites in the dipole regime. We finally outline the technical challenges for future use of NV-based magnetic microscopy in paleomagnetic research.

\paragraph{}% experimental system and measurement protocol
The optical and magnetic imaging was performed using an NV wide-field magnetic microscope setup depicted schematically in Fig. \ref{fig: systemSchematic} (a). To measure the magnetic field we performed continuous-wave optically detected magnetic resonance (CW ODMR) measurements. In this method, The NV spin is optically polarized and interrogated using a green laser at $ 532$ \si{nm}. Concurrently, it is coherently driven by a frequency swept microwave (MW) field, created by a MW antenna located near the diamond surface. The signal is read-out optically via spin-state-dependent fluorescence in a broad range of $ 600-800$ \si{nm}, which is imaged on a camera. Magnetic field projection along the NV orientation is extracted based on the ODMR measured transition frequencies between $m_s=0 \rightarrow +1$ and $m_s=0 \rightarrow -1$ (Fig. \ref{fig: systemSchematic} (d,c)) denoted as $\Delta_{0,+1}$, $\Delta_{0,-1}$ and the Zeeman splitting, which can be written as
\begin{equation}\label{eq: Zeeman}
\begin{aligned}
\Delta_{0,+1}-\Delta_{0,-1} = 2 \gamma B_\parallel
\end{aligned}
\end{equation}
Where $\gamma=28$ \si{GHz/Tesla} represents the electronic gyromagnetic ratio and $B_{\parallel}$ is the external magnetic field along the NV center symmetry axis. 
%The NV spin resonance frequencies are Zeeman-shifted in the presence of a local external magnetic field oriented along the NV center's symmetry axis. Quantitative magnetic field information is extracted based on the measured transition frequency resonances from the ODMR spectrum. [see Fig. \ref{fig: systemSchematic} (d)]. \hl{something is missing here...}

Wide field Magnetic imaging is achieved using a high density ($ \sim 1.3 \times 10^{11}$ \si{NV/cm^2 }) NV ensemble in a thin implanted layer located  $ \sim 10 - 15 $ \si{nm} beneath the surface of the diamond chip (Element 6, $ 4.5 \times 4.5 \times 0.3$ \si{mm} electronic grade diamond, nitrogen implanted with a dose of $2 \times 10^{13}$ \si{N/cm^{2}} and $ 10$ \si{keV} energy). This NV layer is excited over a large field-of-view of $ 35 \times 35$ \si{\mu m}, and the resulting fluorescence is imaged onto a sCMOS camera (Andor Neo) using a high-numerical-aperture (0.9) air objective. Using a single-crystal bulk diamond enables us to measure the magnetic field along each of the four NV center orientations (tetrahedral symmetry) separately, resulting in vectorial magnetic field information. We demonstrate optical imaging of 2D DC magnetic field patterns with $ 350$ \si{nm} spatial resolution and magnetic sensitivity of $ 6 $ \si{\mu T/\sqrt{Hz}} per pixel (calibrated through measurements as depicted in Fig. \ref{fig: systemSchematic} (d)).   

\paragraph{} % data management
The measurement protocol is based on the following procedure. The frequency of the microwave synthesizer was swept around the resonance transition frequencies. For each frequency step, a fluorescence image was taken. The images were appended in the computer memory to form a 3D volume of data, where each element contains the value of the fluorescence for each camera's pixel and for each frequency of the sweep. This volume of data can also be considered as an image where each pixel contains an ODMR spectrum. The ODMR spectrum for each pixel was fitted to a multi-Lorentzian function (8 dips, for 2 resonance transition and 4 NV orientations) (Fig. \ref{fig: systemSchematic} (d)). The algorithm require initial set of parameters for the first fitted pixel. Then the entire image was fitted gradually, taking the results of the already fitted neighboring pixels as an input for the following one. Finally, the resonance positions extracted from the fit are used to reconstruct the magnetic field vector.
%In the results presented in this paper we used only the resonance positions of three of the four NV orientations in order to reduce the measurement time
Calculation of the magnetic field in the lab frame was done by transforming the tetrahedral directions into Cartesian coordinates as defined in Fig. \ref{fig: systemSchematic} (b) using the following relations 
\begin{equation}\label{eq: magnetic field component}
\begin{aligned}
B_x=\frac{-\sqrt{3}}{2}(B_{o2}+B_{o3})\\
B_y=\frac{\sqrt{3}}{2}(B_{o1}+B_{o3})\\
B_z=\frac{-\sqrt{3}}{2}(B_{o1}+B_{o2})
\end{aligned}
\end{equation}
Where, $ B_{oi} $ is the magnetic field along the $i$th NV orientation. To avoid overdetermined coordinate systems we perform a calibration by applying an additional known static magnetic field along spanning set directions. Using the equations in \ref{eq: magnetic field component} we could match each resonance dip in the ODMR spectrum to its NV orientation.  

\paragraph{} % rock measurement protocol
Paleomagnetic imaging measurements were performed by placing a polished rock sample (roughly $ 2 \times 2 \times 1.5$ \si{mm}) on top of the diamond's surface containing the NV center layer (Fig. \ref{fig: systemSchematic} (a)). 
The material we analyze was created by controlled solidification of silicate melt, simulating natural process of magma cooling in volcanic rocks. Details on sample preparation can be found in \cite{Shaar2011,Shaar2010}. The ferromagnetic phase in the sample consists of dendrites of magnetite containing minor amounts of Mn, Mg, and traces of Al, typical to volcanic rocks, homogeneously spread in a paramagnetic silicate matrix. The size, geometry, and composition of the magnetite dendrites allows investigation of domain wall behavior in a plane parallel to a known crystallographic axis. This advantage has been used in previous work by \citet{Shaar2013}, who carried out Magnetic Force Microscopy investigation on this sample. To reduce sample topography and minimize stress effects distorting domain walls, the samples were polished down to a surface roughness of $0.02 $ \si{\mu m} using colloidal silica. We imaged the sample at two different states subsequent to the following magnetic treatment: 1) complete demagnetization state after inducing Alternating Field (AF) with a peak field of 100 mT, and 2) Isothermal remanent magnetization (IRM) after inducing a pulse of of 1.5 T. The IRM field was induced in a direction nearly perpendicular to the polished surface of the sample. Yet, accurate measurement of the angle between the IRM field and the sample surface could not be precisely determined due to technical instrumental limitations. The AF treatment allowed us to investigate domain wall configurations in a "ground state" magnetization scheme, in which domains were arranged spontaneously in the absence of an ambient magnetic field. In contrast, the grain magnetization due to the strong IRM field is maximized, and the latter is used to investigate the potential of NV microscopy as a high sensitivity magnetometer. 

Placement of the polished sample on the polished diamond surface resulted in a stand-off distance of a few microns, which is measured optically (by imaging the diamond surface vs. the sample surface) for each configuration. This stand-off distance translates into a spatial convolution of the measured magnetic signal on the NVs, and therefore into an effectively reduced spatial resolution (on the scale of the stand-off distance), although the inherent optical resolution limit is 350 nm (as stated above). We apply constant magnetic field, so that the magnetic field projection is different along each orientation. In this manner, all eight ODMR resonance dips are resolved (Fig. \ref{fig: systemSchematic} (d)), and measuring three orientations from a single ODMR sweep allows us to extract full quantitative vectorial magnetic field information. We clearly identified a single magnetite dendrite using the same optical path of the fluorescence imaging microscope, and performed the ODMR measurement. Last, we performed a reference measurement excluding the rock sample, which then we used to vectorially subtract from the actual measurement. This resulted in a quantitative magnetic field produced by the rock bulk and the magnetite dendrite solely.

\begin{center} 
\begin{figure}[H]	
	\includegraphics[width=1\columnwidth]{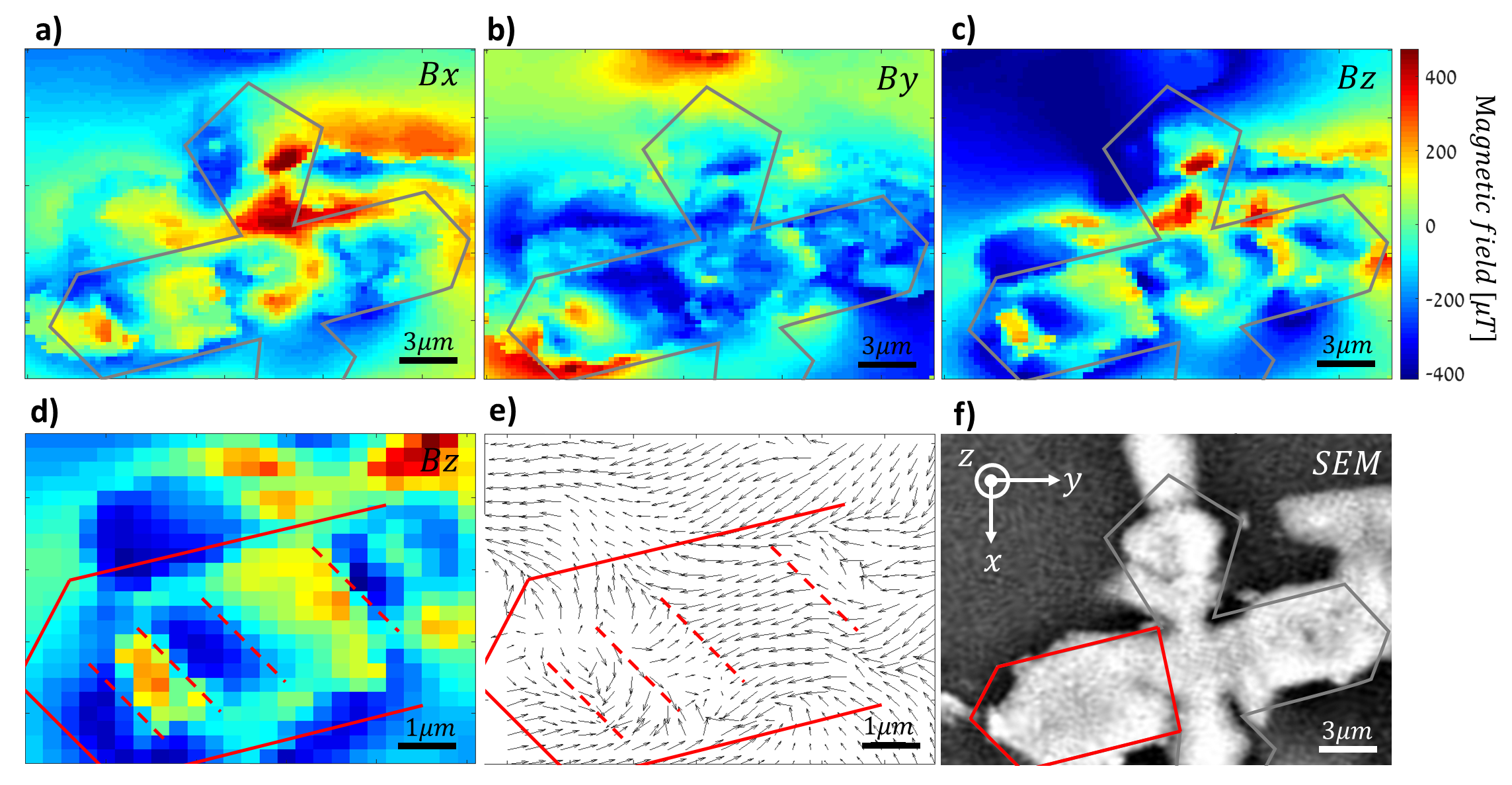} %fig3a.eps
	\caption{Magnetic field of a single magnetite dendrite in the multidomain regime after demagnetization (AF). (a)-(c) Measured magnetic field projections along the x axis (a), y axis (b) and z axis (c) within the same field-of-view. (d) Magnetic field magnification of the left branch of the dendrite along the z direction. (e) Magnetic field projection in the x-y plane of the same left branch of the dendrite as in (d). Red dashed lines represent magnetic domain walls. (f) Scanning Electron Microscope (SEM) image of the dendrite. The gray and red solid lines approximate the location of the magnetite dendrite based on the optical image received from our Wide-field magnetic imaging microscope.}
	\label{fig: magneticImagingFlakev5}
\end{figure}
\end{center}

\paragraph{} % MD results
First, we present in Fig. \ref{fig: magneticImagingFlakev5} the results of the demagnetized sample at a measurement distance of $ 3$ \si{\mu m} in the presence of a constant bias magnetic field of $ 5 $ \si{mT}  along the (-0.8,-0.5,-0.4) direction. The results show the quantitative vectorial magnetic field of a single magnetite dendrite. The branches of magnetite dendrite occur as rod-like extensions growing along magnetite $\langle 100 \rangle$ directions. The size of each individual branch of the dendrite correspond to a MD regime \cite{Muxworthy2006}. The images of the magnetic field show domains with anti-parallel magnetization separated by sharp walls along the dendrite branch. The steep change in magnetic field orientation over a relatively short distance corresponds to such multi-domain behavior, and is captured by our measurement scheme (although spatial resolution is somewhat limited due to the stand-off distance between diamond and sample). The geometry, relative size, pattern, and directions of the domains are similar to those observed using MFM by \cite{Shaar2013}, who also showed that the preferred direction of the walls is along the $\langle 111 \rangle$ crystallographic plane, explaining the repetitive pattern in the left branch shown in Fig. \ref{fig: magneticImagingFlakev5}(d).

% MD results image

\begin{center}
\begin{figure}[tbh]	
	\includegraphics[width=1\columnwidth]{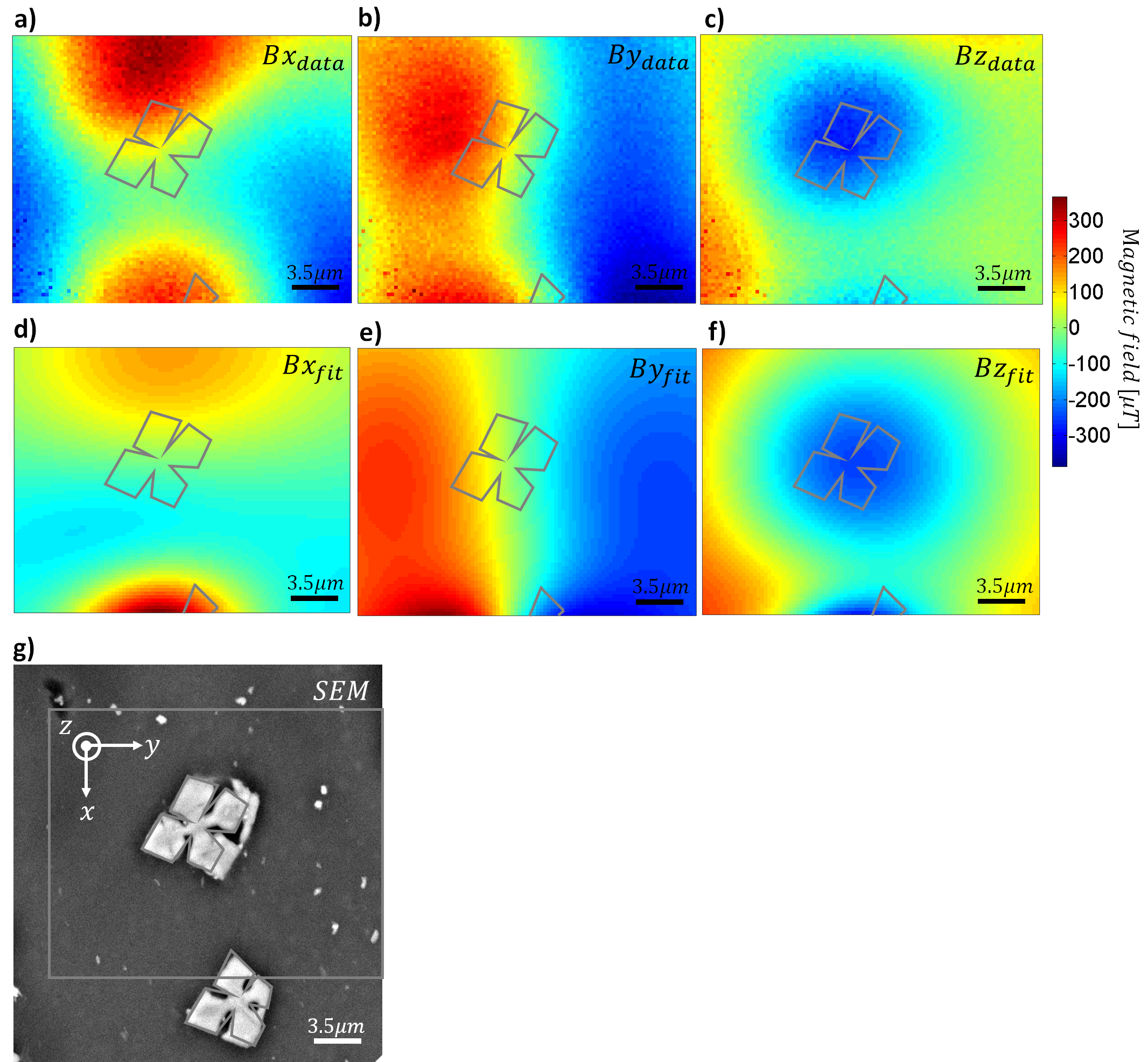} %fig3a.eps
	\caption{Magnetic field of magnetite dendrites in the dipole-dominated regime. (a)-(c), Measured magnetic field projections along the x axis (a), y axis (b) and z axis (c) within the same field-of-view. (d)-(f), Simulated magnetic field projections along the x axis (d), y axis (e) and z axis (f). (g), Scanning Electron Microscope (SEM) image of the dendrite (the magnetic images' field of view is denoted by the gray square). The gray solid lines approximate the location of the magnetite dendrite based on the optical image received from our Wide-field magnetic imaging microscope.}
	\label{fig: magneticDipole}
\end{figure}
\end{center}

\paragraph{} % dipole results
Second, we present the results of a different, smaller particle after inducing IRM field in the -z direction. The measurements distance to the sample is $ 10$ \si{\mu m} in the presence of a constant bias magnetic field of $ 2$ \si{mT} along the (0.3,0.7,-0.6) direction. This distance from the sample is far enough to be in a dipole regime, i.e. observing magnetic field lines similar to those of a simple dipole. This distance is on the other hand close enough to result in relatively strong field measurements. Here we used the dipole approximation to calculate the effective magnetic moment of the particle. The magnetic moment calculation was performed based on the full vectorial magnetic images, the sensor-to-sample distance and the position of the dendrite in the x-y plane. A nonlinear fit was performed to a simple magnetic dipole, leaving the dendrite depth and the vectorial magnetic moment as free parameters. 
The results in Fig. \ref{fig: magneticDipole} show the quantitative vectorial magnetic field of two magnetite dendrites and the corresponding fit of the magnetic field assuming tilted dipoles centered under the grains. The  magnetic moment of the dendrite in the upper part of Fig. 3 is $ \sim 1.2 \times 10^{-11}$ \si{Am^2} along the (0,-0.6,-0.8) direction. The direction of the dipole moment is in agreement with the direction of the applied field. Assuming a pure magnetite (saturation magnetization, Ms, of $4.8 \times 10^{5}$  \si{{A/m}}) dendrite built from six octahedra with edge length of $3.2$ \si{\mu m} the saturation magnetization of a half dendrite, as the one shown in Fig. 3, is $  2.2 \times 10^{-11}$ \si{Am^2}. Following \citet{Tauxe2002,Williams2010} it is reasonable to assume that the IRM is about half of the saturation magnetization, which result is magnetic moment of $1.1 \times 10^{-11}$  \si{A m^{2}}, very similar to the best-fitting dipole in Fig. \ref{fig: magneticDipole} (d-f).

% dipole measurement image

In both measurements, to separate the paramagnetic magnetization of the silicate matrix from the ferromagnetic effect, we measured the magnetic field of the matrix away from the magnetite dendrite (this field is not zero due to the finite size of the sample, inhomogeneities of the material, and contributions from ferromagnetic grains). A vectorial subtraction of this measurement from the dendrite measurement resulted in an approximation of the magnetic field of the magnetite alone. In the first result (Fig. \ref{fig: magneticImagingFlakev5}), the magnetic field of the rock matrix was in a direction close to constant bias magnetic field and a magnitude of $ 0.2$ \si{mT} . In the second result (Fig. \ref{fig: magneticDipole}), the magnetic field magnitude of the rock matrix was $ 1$ \si{mT} along the (0.6,0.6,-0.5) direction. We expect that the discrepancy between the orientation of this field and the external field stem from field inhomogeneities and ferromagnetic effects. As detailed below, more precise measurements in the future will utilize a 3D Helmholtz coil configuration, affording better field homogeneity and a clearer separation between paramagnetic and ferromagnetic effects.

%Second, we present the results of the sample after TRM acquisition at multiple measurement distances in the presence of a constant bias magnetic field of $ 2 \, mTesla $. The results in Fig. \ref{fig: magneticImagingFlakeVsDistanceV3} show the quantitative vectorial magnetic field of two magnetite dendrites. To receive more information about the magnetic field behavior, we performed this measurement in three distances, $ 7.5$,  $12 $ and $ 21$ $ \mu$ m .  Interestingly, the field image cannot be easily correlated to the ferromagnetic particles. That is, the flux is not concentrated above the samples. ...
%\hl{I need to Think about it. Not sure yet how to explian how we get magnetixation in the -z direction... }

%\begin{center}
%\begin{figure}[H]	
%	\includegraphics[width=1\columnwidth]{magneticImagingFlakeVsDistanceV3.png} %fig3a.eps
%	\caption{Magnetic field of magnetite dendrites in the dipole-dominated regime at different distances. (a)–(i), Measured magnetic field projections in a sensor-to-sample distances of $ 7.5 \, \mu m $ (d-f), $ 12 \, \mu m $ (g-i) and $ 21\, \mu m $ (a-b) along the Cartesian coordinates denoted in the low right corner.}
%	\label{fig: magneticImagingFlakeVsDistanceV3}
%\end{figure}
%\end{center} 

%\hl{Third, demagnetization results and magnetic corrosivity}
%%%%%%%%%%%%%%%%%%%%%%%%%%%%%%%%%%%%%%%%%%%%%%%%%%%%%%
%% Summary & Conclusions
%%%%%%%%%%%%%%%%%%%%%%%%%%%%%%%%%%%%%%%%%%%%%%%%%%%%%%

%\paragraph{} % current capabilities and measurements

\paragraph{} % future capabilities and applications 

\paragraph{} 
The results presented here demonstrate the capabilities and the future potential of NV microscopy in measuring the magnetic signals of individual micrometer scale grains. Our case study includes optical and magnetic imaging of geological magnetite dendrites from a distance of $ 3-10$ \si{\mu m}, with $ 350$ \si{nm} spatial resolution, magnetic sensitivity of $6$ \si{\mu T/\sqrt{Hz}} and a field of view of $35$ \si{\mu m}. 

In the demonstrated measurement modality we were limited by our control over the  $ 2 $ \si{mT} bias magnetic field. The bias field exerts a paramagnetic effect on silicates, and can distort magnetic domain configuration in grains with low coercive field. This problem can be solved by adding 3D Helmholz coils \cite{Fu2014}, which should allow precise control over the vectorial bias field in future work. We expect to be able to perform measurements with bias magnetic field smaller by approximately an order of magnitude ($\sim 0.5$ \si{mT}), as well as to reverse the direction of the bias field in order to separate paramagnetic and ferromagnetic effects.  

\paragraph{} 
In this paper we demonstrated two potentially useful applications of NV microscopy in paleomagnetism. The first is the direct measurement of the full magnetic vector created by magnetic domains (Fig. \ref{fig: magneticImagingFlakev5}). This application requires measurements as close as possible to the sample surface, in order to increase the spatial resolution in detecting fine-scale domain configurations. The second application is the calculation of weak magnetic moments of individual grains, the size of which is in the range of a few micrometers. This capability can potentially open a new path for paleomagnetic analyses of weakly magnetized individual ferromagnetic grains, such as minerals hosted in meteorites \cite{Fu2014} and the oldest rocks on Earth \cite{Tarduno2015}. Calculations of such small-scale moments are currently employed using low-temperature scanning SQUID devices that measure the z component of the vector \cite{Lima2016}. Adopting the approach of \cite{Lima2016}, using the three components of the field vector at room temperature offers promising potential for future applications. Altogether, these two applications provide significant supplements to the possibilities of NV microscopy in geological research, complementing the paleomagnetic applications first demonstrated by \citet{Fu2014,Fu2017}.

Finally, we wish to highlight some technical challenges in application of NV magnetometry in paleomagnetism. First, the bias field should be minimized in order to reduce distortion of the paleomagnetic signal by reorganization of domain walls as a response to external field. Then, one should make sure that the coercive field of the investigated samples, i.e. the field required to trigger domain-wall reorganization, is lower than the NV bias field. Second, we stress the necessity of generating a uniform bias field using a controlled 3-axis Helmholtz coil configuration \cite{Fu2014,Fu2017}. This will allow better control of paramagnetic effect, and will enable subtraction of the paramagnetic signal from the ferromagnetic one by carrying out two measurements with antiparallel fields. Third, to support the use of NV magnetometry in standard paleomagnetic treatments, such as progressive AF demagnetizations (Fig. \ref{fig: magneticImagingFlakev5}), the design of a specialized sample holder is required to allow reproducible placement of the sample with respect to the diamond detector. Fourth, the standoff distance between the diamond and the rock surface should be adjustable such that the measurement regime (dipole-dominated or multi-domain) could be determined.  

This work has been supported in part by the CIFAR-Azrieli global scholars program, the Israel Science Foundation (grant No. 750/14), the Ministry of Science and Technology, Israel, and the KLA fellowship.

%We showed direct observations of MD magnetization behavior in a single magnetite dendrite, which acquire its magnetization by the $ 5 \, mTesla $ biased magnetic field. In addition, we were able to extract magnetic dipole moment from a single dendrite magnetized by the $ 60 \, \mu Tesla $ TRM process. \hl{anybody have something to add here?}
\paragraph{} % biased magnetic field
%All of our TRM and SRM measurement results were affected by the $ 2 \, m Tesla $ biased magnetic field. Therefore, in addition to the magnetization signal acquired by the TRM and the SRM process, we measured a signal created by the paramagnetic and the ferromagnetic effects. Currently, we cannot eliminate or know these effects at the dendrite level due to the the lack of theory on MD behavior and the lack of knowledge about the dendrites corrosivity. We know from our bulk measurements that our NV measurement procedure affects the bulk magnetization, hence we cannot say whether or not it affected the specific dendrites we measured.

\bibliography{nvbibliography}

\end{document}